\documentclass[footinbib,aps,pra,reprint,superscriptaddress]{revtex4-2}
\usepackage{amsmath,amssymb,braket,upgreek,bm,verbatim}
\usepackage{graphicx,tabularx} 
\usepackage{float} %
\usepackage{blindtext}
\usepackage{xcolor}

\usepackage[absolute]{textpos}
\usepackage{xcolor}
\usepackage{multirow}
\usepackage[colorlinks=true,bookmarks=false,citecolor=blue,urlcolor=blue]{hyperref}

\newcolumntype{Y}{>{\centering\arraybackslash}X}
\def\cF{\mathcal{F}}

\begin{document}
\title{Advanced Architectures for High-Performance Quantum Networking}

\author{Muneer Alshowkan}
\email{alshowkanm@ornl.gov}
\affiliation{Computational Sciences and Engineering Division, Oak Ridge National Laboratory, Oak Ridge, Tennessee 37831, USA}

\author{Philip G. Evans}
\affiliation{Computational Sciences and Engineering Division, Oak Ridge National Laboratory, Oak Ridge, Tennessee 37831, USA}

\author{Brian P. Williams}
\affiliation{Computational Sciences and Engineering Division, Oak Ridge National Laboratory, Oak Ridge, Tennessee 37831, USA}

\author{Nageswara~S.~V. Rao}
\affiliation{Computational Sciences and Engineering Division, Oak Ridge National Laboratory, Oak Ridge, Tennessee 37831, USA}

\author{Claire~E. Marvinney}
\affiliation{Computational Sciences and Engineering Division, Oak Ridge National Laboratory, Oak Ridge, Tennessee 37831, USA}

\author{Yun-Yi Pai}
\author{Benjamin~J. Lawrie}
\affiliation{Materials Science and Technology Division,\\ Oak Ridge National Laboratory, Oak Ridge, Tennessee 37831, USA}

\author{Nicholas~A. Peters}
\affiliation{Computational Sciences and Engineering Division, Oak Ridge National Laboratory, Oak Ridge, Tennessee 37831, USA}

\author{Joseph~M. Lukens}
\email{lukensjm@ornl.gov}
\affiliation{Computational Sciences and Engineering Division, Oak Ridge National Laboratory, Oak Ridge, Tennessee 37831, USA}

\date{\today}

\begin{abstract}
\label{Abstract}
As practical quantum networks prepare to serve an ever-expanding number of nodes, there has grown a need for advanced auxiliary classical systems that support the quantum protocols and maintain compatibility with the existing fiber-optic infrastructure.
We propose and demonstrate a quantum local area network design that addresses current deployment limitations in timing and security in a scalable fashion using commercial off-the-shelf components. We %
employ White Rabbit switches to synchronize three remote nodes with ultra-low timing jitter,  
significantly increasing the fidelities of the distributed entangled states over previous work with Global Positioning System clocks. Second, using a parallel quantum key distribution channel, we secure the classical communications needed for instrument control and data management. 
In this way, the conventional network which manages our entanglement network is secured using keys generated via an underlying quantum key distribution layer, preserving the integrity of the supporting systems and the relevant data  in a future-proof fashion.
\end{abstract}
\maketitle

\begin{textblock}{13.3}(1.4,15)
\noindent\fontsize{7}{7}\selectfont \textcolor{black!30}{This manuscript has been co-authored by UT-Battelle, LLC, under contract DE-AC05-00OR22725 with the US Department of Energy (DOE). The US government retains and the publisher, by accepting the article for publication, acknowledges that the US government retains a nonexclusive, paid-up, irrevocable, worldwide license to publish or reproduce the published form of this manuscript, or allow others to do so, for US government purposes. DOE will provide public access to these results of federally sponsored research in accordance with the DOE Public Access Plan (http://energy.gov/downloads/doe-public-access-plan).}
\end{textblock}

\section{Introduction}
\label{Introduction}
Quantum networks are essential for the optimal utilization of fundamental quantum resources. %
Advances in quantum information science including distributed computing~\cite{Cirac1999distributed}, enhanced sensing~\cite{Bollinger1996optimal,Giovannetti2004quantum}, secure communications~\cite{Bennett2014quantum}, and blind computing~\cite{Broadbent2009universal} will depend on 
communication between quantum devices over
quantum local area networks (QLANs), quantum metropolitan area networks (QMANs), quantum wide area networks (QWANs), and
the future ``quantum internet'' (QI)~\cite{Kimble2008the, Wehner2018quantum}.
Benefits of entanglement distribution over short- and medium-range networks such as QLANs and QMANs include distributed computing and secure communications, %
whereas entanglement distribution over long-range networks such as QWANs and the QI will enable long-baseline interferometry~\cite{Gottesman2012longer} and enhanced sensing~\cite{Zhuang2018distributed} on a global scale.

Early methods to allocate entanglement to many users %
include probabilistic distribution using passive beam splitters~\cite{Townsend1997quantum} and reconfigurable spatial splitting using optical switches~\cite{Toliver2003experimental}, often combined with  fixed wavelength division multiplexing~\cite{Peters2009dense, Chapuran2009optical, Herbauts2013demonstration, Laudenbach2020flexible}.
Fully connected quantum key distribution (QKD) networks based on nested dense wavelength-division multiplexers (DWDMs) have been explored in Refs.~\cite{Wengerowsky2018entanglement, Joshi2020trusted}, yet the complex network of DWDMs
results in high variability in loss and makes the architecture challenging to scale. An alternative approach to fully connected wavelength-multiplexed networks uses a flex-grid wavelength-selective switch (WSS), which enables reconfigurability in \emph{both} the spatial connections \emph{and} channel bandwidths~\cite{Lingaraju2021adaptive, Appas2021}---a combination lacking in previous approaches. %
In a broader context, these networks constitute the basic architectural building blocks (namely, at 0-order of recursion) of the Quantum Recursive Network Architecture (QRNA) \cite{VanMeter2014book}, conceived to be composed of (future) quantum repeaters and routers.
Recently, we demonstrated flex-grid entanglement distribution in a deployed network for the first time, connecting nodes in three campus buildings that were time-synchronized by the Global Positioning System (GPS)~\cite{Alshowkan2021qlan}. %
Although the nanosecond-scale jitter of GPS synchronization permitted clear verification of entanglement between distributed sites, we noted that the overall fidelity was strongly reduced
by the background noise included within the jitter-limited coincidence window, thereby highlighting the crucial importance of classical clock distribution technologies for quantum network performance. %

In addition to timing synchronization, quantum networks depend heavily on classical communications for instrument control and data collection as well. The security of these classical communications is critical, and it is desirable to establish it in a technology-independent way, i.e.,  regardless of quantum or conventional computing resources leveraged against it. 
This consideration moves beyond the concerns of a QKD network, where the quantum resources are utilized to secure some classical application; here, the question focuses on securing the results of quantum experiments themselves from eavesdropping or modification. Securing the classical channel is essential for the integrity of the entanglement distribution process as well as any protocols utilizing the distributed entangled resources.

In this paper, we propose and demonstrate an upgraded QLAN that addresses both of these limitations in a scalable design with commercial off-the-shelf components, improving the quality of entanglement distribution and securing the classical communications needed to support the network. First, we utilize White Rabbit (WR) clock synchronization between the distributed network nodes, %
designing our field-programmable gate array (FPGA) time-to-digital converter (TDC) at each location to achieve timing resolution at the limits of available hardware, significantly improving the fidelity of entanglement distribution over our previous GPS-based demonstration. %
Additionally, we secure the network's classical communications for instrument control and data management with commercial firewalls that utilize quantum keys generated in a parallel channel, preserving the integrity of all data over a public network. %
To our knowledge, these results represent the first use of White Rabbit in a deployed quantum network and the first general-purpose entanglement distribution network with all ancillary classical communications secured by QKD. Our findings demonstrate the promise of scalable quantum network synchronization and security with plug-and-play systems that integrate seamlessly within existing fiber-optic and ethernet ecosystems.

\begin{figure}[t!]
\includegraphics[width=\columnwidth]{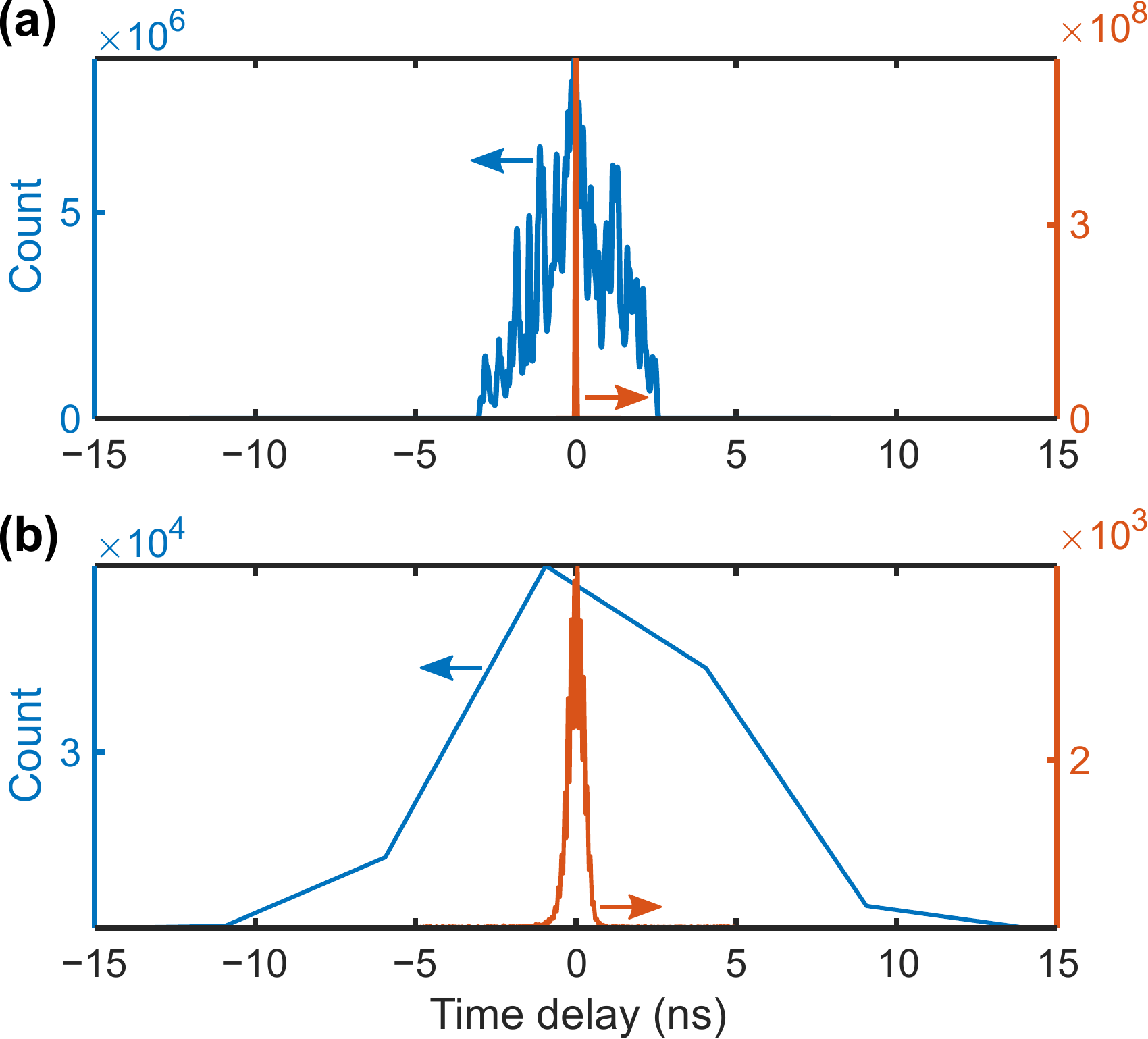}
\caption{(a)~Histogram of relative delays between 10~MHz clocks from two GPS receivers (blue) and two WR receivers (orange), integrated over 30~minutes and plotted with 1~ps binning. (b)~Distribution of all coincidence counts collected during a 37-step waveplate scan between Alice--Charlie using GPS (blue) and WR (orange) clock synchronization with 60~s and 30~s integration time, respectively. In each plot, arrows point to relevant y-axis for each curve.}
\label{fig_jitter_multi}
\vspace{-0.05in}
\end{figure}

\section{Timing Synchronization}
The exciting potential of quantum networks can be attributed to their ability to establish and exploit entanglement between spatially distributed nodes. As measurements verifying and operating on entangled parties must be tightly synchronized, low-jitter clock distribution represents a fundamental prerequisite of successful quantum networks. In many previous quantum communications experiments~\cite{Hughes2005quantum, Chapuran2009optical}, optical synchronization pulses that offer extremely low jitter have been the method of choice. Yet such systems generally require significant engineering as well as specialized, often expensive components, making it unclear how to scale optical synchronization methods to arbitrary numbers of nodes in a practical setting.
On the other hand, GPS offers a universal and cost-effective solution for time synchronization with nanosecond-level precision; for this reason, it was our choice in previous work~\cite{Alshowkan2021qlan}. Yet GPS fails to reach the picosecond-scales desired in many photon coincidence counting contexts. 

Interestingly, the need in quantum networks for precise time tagging of distributed events is shared by large-scale particle physics experiments, so it is perhaps unsurprising that the particle-accelerator-inspired technology of WR would prove well-suited for quantum networking as well.
Recently incorporated into the IEEE1588-2019 Precision Time Protocol (PTP) standard~\cite{IEEE1588-2019}, WR utilizes synchronous ethernet and phase detection to achieve relative clock jitters on the order of a few picoseconds~\cite{Lipinski2011white, Rizzi2018}. Indeed, with such precision, WR has been employed to synchronize optical cameras for characterizing an entangled photon source~\cite{Nomerotski2020Spatial}.
Fortunately, WR is built on the concept of open-source hardware, firmware, and software, so WR equipment can either be purchased from commercial vendors who offer it under a CERN open hardware licence or built in-house from available devices. Therefore, users can utilize and build on top of solutions well established by both open-source and commercial techniques. 
Notably, both GPS and WR systems can directly output 1~Hz and 10~MHz clocks, making WR drop-in compatible with existing GPS-based time synchronization modules. To compare the performance of the 10~MHz clocks from GPS and WR, we measured the relative delay between a pair of GPS receivers (Trimble Thunderbolt E) and a pair of WR nodes (Seven Solutions WR-LEN) connected to the WR switch (Seven Solution EQP-WRS-LJ-01)---all nodes independent but located in the same laboratory for this test---using a fast TDC (Swabian Instruments Time Tagger Ultra). The distributions of relative delays between each pair of receivers, recorded over 30~minutes and shown in Fig.~\ref{fig_jitter_multi}(a), have standard deviations of 1.21~ns and 12.9~ps for GPS and WR, respectively~\cite{endnote1}. The compatibility of this design offers scalable time synchronization and time tagging devices with reasonable cost and complexity via commercial components.

\begin{figure*}[tb!]
	\centering
    \includegraphics[width=\textwidth]{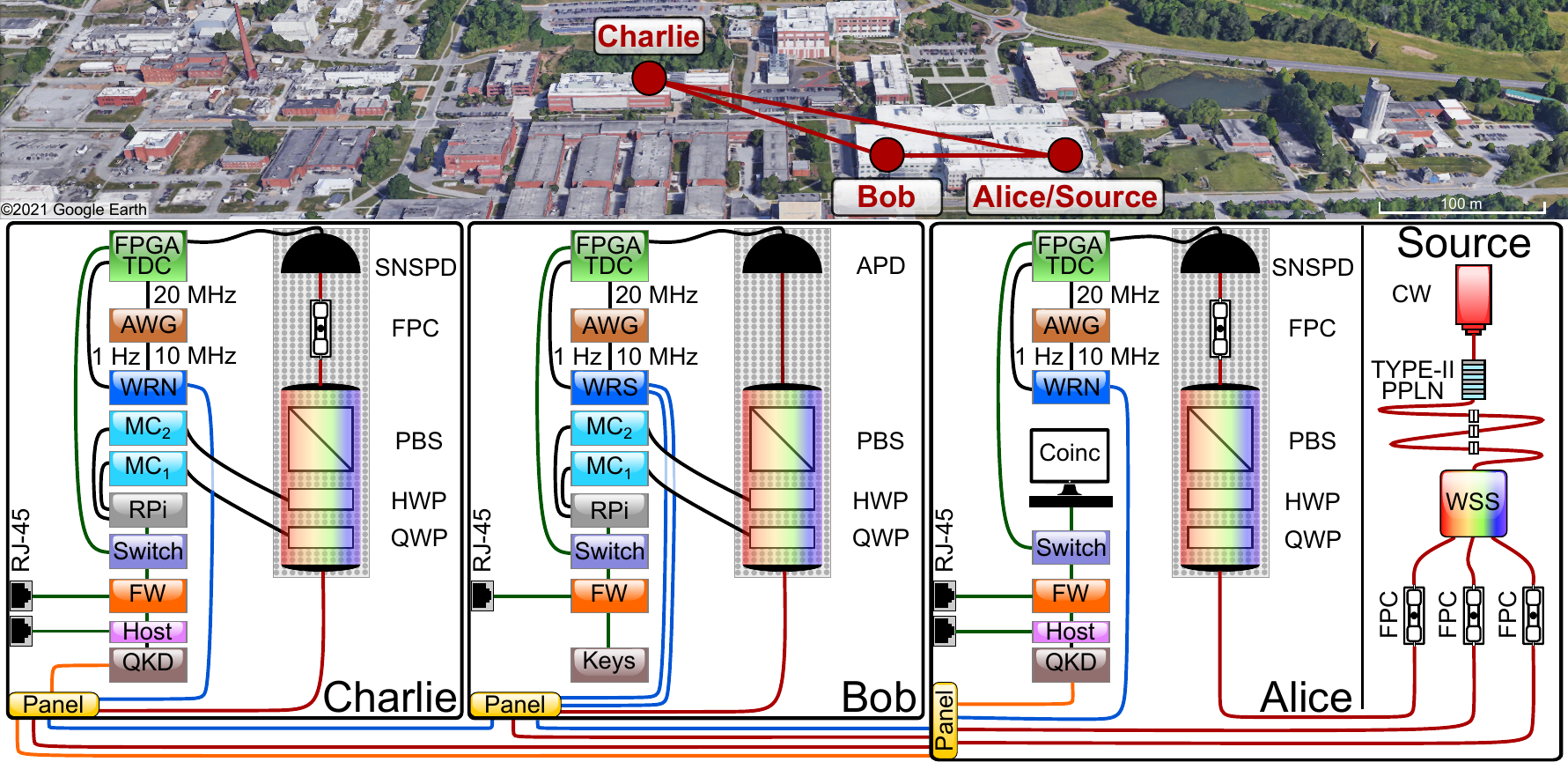}
	\caption{Quantum local area network (QLAN) setup. The receiver configurations at each node are shown as insets. 
	APD: avalanche photodiode. 
	AWG: arbitrary waveform generator. 
	BS: beamsplitter. 
	Coinc: computer system for data collection and coincidence counting. 
	CW: continuous-wave laser. 
	FPC: fiber polarization controller. 
	FPGA: field-programmable gate array. 
	FW: network firewall.  
	Host: QKD host machine. 
	HWP: half-wave plate. 
	MC: motion controller. 
	Panel: fiber-optic patch panel. 
	PBS: polarizing beamsplitter.
	Keys: pre-generated quantum keys.
	PPLN: periodically poled lithium niobate. 
	QKD: quantum key distribution module. 
	QWP: quarter-wave plate. 
	RJ--45: Ethernet connection to the campus network. 
	RPi: Raspberry Pi microprocessor board (to control MCs). 
	SNSPD: superconducting nanowire single-photon detector. 
	Switch: ethernet switch. 
	WSS: wavelength-selective switch. 
	WRN: White Rabbit node. 
	WRS: White Rabbit switch. 
	Red lines: quantum source signals (optical). 
	Orange lines: QKD signals (optical). 
	Blue lines: WR signals (optical). 
	Green lines: ethernet signals (electrical). 
	Black lines: other timing and control signals (electrical).}
	\label{fig_setup_map}
	\vspace{-0.05in}
\end{figure*}

Previously, we designed our FPGA TDCs to log detection events in 5~ns bins according to an internal 200~MHz clock~\cite{Alshowkan2021qlan}---reasonable for GPS synchronization but entirely ill-suited to the timing resolution possible with WR.
To reach significantly finer resolutions without any new hardware, a tapped delay line (TDL) can be constructed on the FPGA, such as that formed by carry4 adders~\cite{Song2006, Jinhong2010fully, Zhao2013design, Jiajun2017low, Adamic2019fast}. To understand the basic principle of operation, consider a long binary adder, where one added is initialized to all ones and the other to all zeros. The logical pulse from a detector is then fed into the first (least-significant) bit position as a carry-in bit of 1, causing the carry-out bit to switch from 0 to 1 and initiating a cascade as the carry-out bits from all positions in the chain switch to 1. Thus, by recording the total number of nonzero carries at the end of each clock cycle, it is possible to back out the original pulse arrival time with deep sub-cycle precision, as set by physical propagation on the FPGA board.

In our network, each distributed node has WR-derived 10~MHz and 1~Hz clocks. While the 1~Hz time pulse is connected directly to the FPGA and counted, the 10~MHz clock is first frequency-doubled %
using an arbitrary waveform generator and then fed into a phase-locked loop (PLL) to derive internally synchronized clocks (the specific FPGA requires PLL input frequencies $>19$~MHz~\cite{Xilinx2018}, precluding a direct lock to 10~MHz). We use the PLL to produce 10~MHz and 200~MHz clocks on-board the FPGA---which is similar to our previous use of GPS synchronization, but now the 200~MHz clock is also locked to the WR signal, rather than free-running. For fine timing within each 5~ns clock period, our TDL of carry4 adders bins detection events with a fundamental bin width of approximately 17.5~ps per bit, as determined by recording the number of bits that are switched within the 5~ns clock period. We note that this bin width is comparable to others reported in the literature~\cite{Zhao2013design, Jiajun2017low, Adamic2019fast}. To compare the improvement of our design, Fig.~\ref{fig_jitter_multi}(b) shows example histograms of coincidence events between Alice and Charlie detected by our previous GPS- and new WR-based designs. These curves include all jitter contributions (clock distribution, detector, and FPGA) and show the different bin sizes used (5~ns and 17.5~ps). From these tests, we were able to select a 1~ns coincidence window for the upgraded QLAN compared to 10~ns before~\cite{Alshowkan2021qlan}, the practical consequence of which is an approximately $10$-fold reduction in accidental coincidences.

\section{Quantum Security}
A quantum network will require conventional network capabilities in the form of a control plane for management and a parallel data plane for classical communication between nodes~\cite{Dahlberg2019link, Alejandro2019engineering, Diego2020demonstration}. In addition, classical data flow supporting quantum networks must be secure. %
To address this gap, we establish a classical control plane to route classical data used for instrument control and data management over the campus networking infrastructure. Then, we key the established control plane with keys from a commercial QKD system (ID Quantique). %
The control plane is constructed with commercial NIST-certified networking firewalls (Palo Alto PA-220) that enable network traffic control and encryption. Each firewall employs the advanced encryption standard (AES) algorithm~\cite{Daemen2000aes} based on cipher block chaining (CBC)~\cite{dworkin2001recommendation} with 256-bit secret keys pulled from the QKD system. The QKD host system has two active interfaces in each location: a private connection with the local firewall to supply the secret keys and another to the campus networking infrastructure for transmitting and receiving the QKD-authenticated classical communications required to establish a new key. In our design, we are free to choose the secret key update intervals. However, special consideration must be given to the type of traffic en route when the secret key changes. While the transmission control protocol (TCP) tracks and expects acknowledgment for packets received---missing an acknowledgment triggers re-transmission of lost packets---the user datagram protocol (UDP) does not offer the same reliable service. Therefore, in order to avoid data loss during a quantum entanglement measurement, we selected a key update interval longer than the total measurement time. %
We have found this general principle of update synchronization adequate to offer compatibility with devices utilizing UDP communications, such as our FGPAs here.

\begin{figure}[t!]
\centering
\includegraphics[width=2.8in]{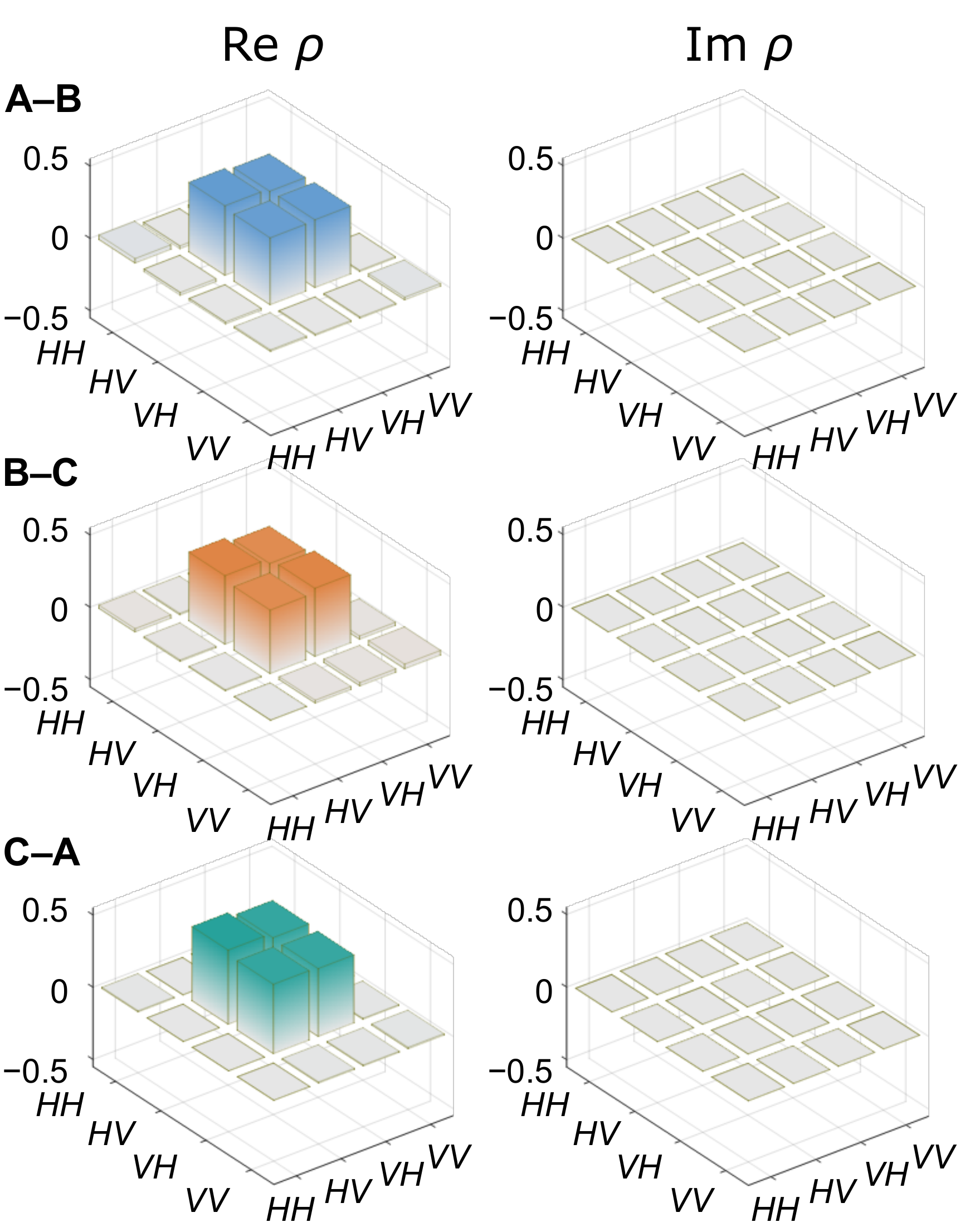}
\caption{Density matrices as estimated by polarization tomography for each pair of users. A--B: Alice--Bob, B--C: Bob--Charlie, C--A: Charlie--Alice.}
\label{fig_density}
\vspace{-0.05in}
\end{figure}

\section{Experiment}
As shown in Fig.~\ref{fig_setup_map}, the experiment is performed in three buildings on the Oak Ridge National Laboratory campus. While the source and Alice share a lab, Bob and Charlie are located in different buildings; each connects to the source's patch panel via fiber path lengths of approximately 250 and 1200 meters, respectively. The description and characterization of the source are covered extensively in Ref.~\cite{Alshowkan2021qlan}. The WR clock distribution switch is located at Bob and connects to Alice and Charlie via separate fiber links. Classical communications are facilitated by a control plane set up over the campus infrastructure and routed between the local networking firewalls. We connect the public interface to the conventional campus network and the private interface to the QKD host system to supply the secret key via a direct physical ethernet connection. 

\begin{figure}[b!] %
\includegraphics[width=\columnwidth]{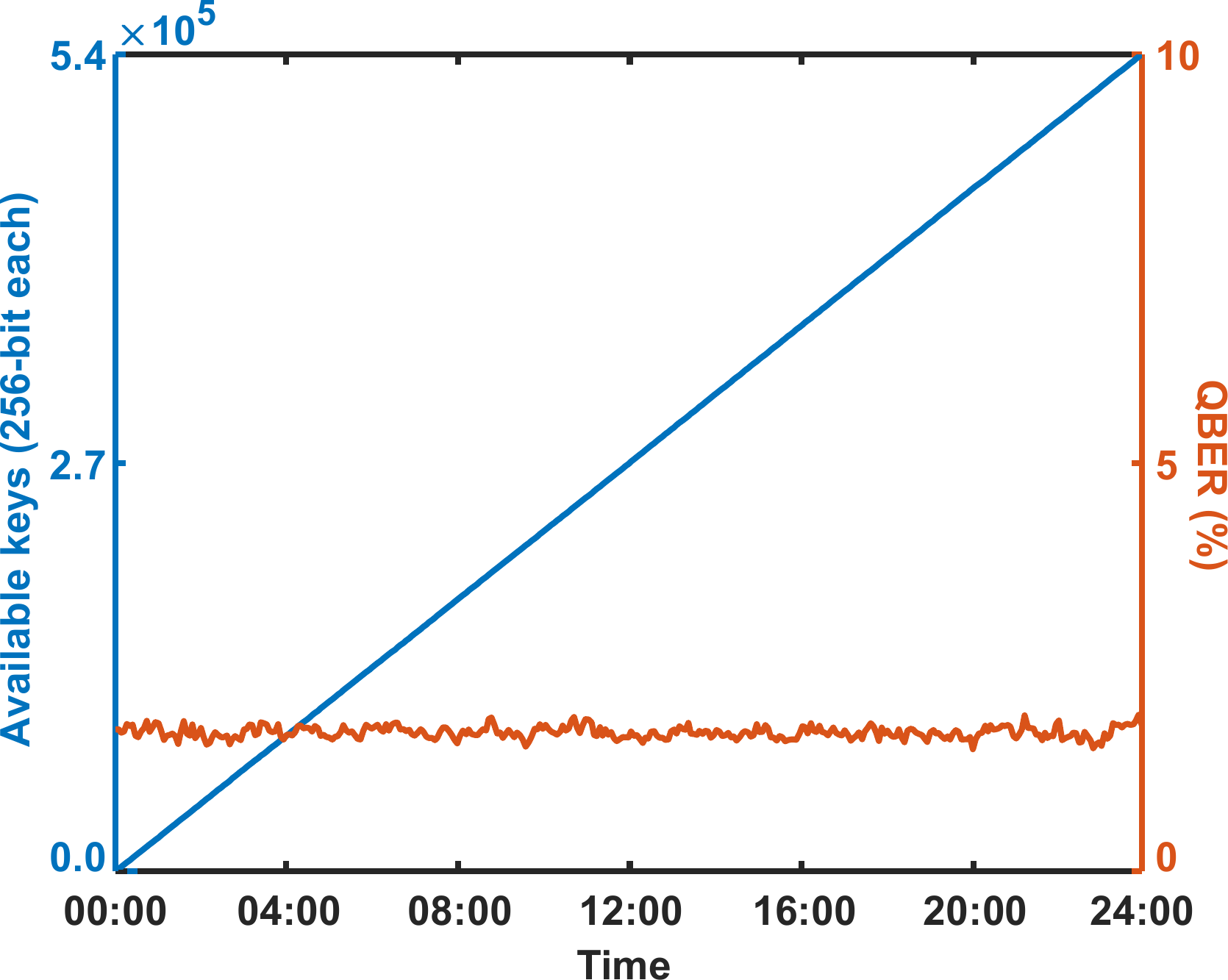}
\caption{Total number of available 256-bit secret keys (blue) and measured QBER (orange) as functions of time. A linear fit to the available keys gives an average key production rate of $6.2781\pm0.0004$~keys/s. }
\label{fig_key_rate_error}
\vspace{-0.05in}
\end{figure}

In this experiment, there are three possible bipartite entanglement links: A--B, B--C, and C--A. However, only one QKD system is available, which we deploy on the C--A link for concurrent key distribution. For A--B and B--C, we allocate to each user a file of pre-generated quantum keys which are then applied to each firewall at regular intervals, emulating the performance of real-time key distribution at all nodes. 
The QKD system employed in this experiment uses a phase-encoded SARG04 protocol~\cite{Scarani2004quantum}. %
It begins with a raw key exchange phase over the quantum channel. Then over the classical channel, it performs sifting, error correction, and privacy amplification. %
Each successful cycle produces a long string of secret bits on the relevant QKD host systems at each location. We developed software running on the QKD hosts to monitor the file for fresh key material, then chunk the long key string into 256-bit strings that the firewall accepts as an input to the AES-256-CBC encryption algorithm. %
Data collected during twenty-four hours of continuous operation show the performance of QKD between Alice and Charlie in Fig.~\ref{fig_key_rate_error} with average secret key rate (SKR) and quantum bit error rate (QBER) of $1620 \pm 150$ bits/s and $1.68 \pm 0.09$ \%, respectively. Since a full tomographic measurement (which includes a waveplate scan to compensate any residual $H/V$ phase~\cite{Alshowkan2021qlan}) takes $\sim$37~minutes, we design the software in each location to update the firewall encryption algorithm with a new secret key every 40~minutes, thus establishing a QKD-secured classical connection between two QLAN nodes for entanglement distribution. %

In the source, a continuous-wave laser set at 779.4~nm pumps a periodically poled lithium niobate crystal to generate spectrally correlated, polarization-entangled photons in the ideal Bell state $| \Psi^+\rangle\propto\ket{HV}\pm\ket{VH}$ via type-II spontaneous parametric downconversion~\cite{Lingaraju2021adaptive,Alshowkan2021qlan,Kaiser2012high}. A WSS receives the generated bandwidth of biphotons and sections it into eight pairs of channels, each entangled in frequency. Each WSS output is connected to a fiber polarization controller for polarization compensation along the path before reaching each user's polarization analysis module. %
The output of the polarization analyzer is directed to an InGaAs avalanche photodiode for Bob and to superconducting nanowire single-photon detectors preceded by fiber polarization controllers to maximize detection efficiency for Alice and Charlie. Taking the previous QLAN results as a benchmark~\cite{Alshowkan2021qlan}, we select a combination of channel allocations corresponding to the highest fidelities previously obtained for each link: Alice and Bob with Ch. 3, Bob and Charlie with Chs. 1--2, and Charlie and Alice with Ch. 8. We measure the quality of the distributed state by performing polarization tomography from data processed with a 1-ns coincidence window and 30~s integration time per point. We use the Bayesian tomography method of Ref.~\cite{Lukens2020practical} to obtain the density matrices shown in Fig.~\ref{fig_density} for each pair of nodes, combining data from measurements in rectilinear and diagonal polarization bases. The fidelities with respect to the ideal Bell state $|\Psi^+\rangle$, logarithmic negativities $E_\mathcal{N}$~\cite{Vidal2002}, and entanglement rates $R_E$~\cite{Alshowkan2021qlan} are shown in Table~\ref{table_result}, with no subtraction of accidentals. 

Under the finer timing resolution, fidelities have improved significantly across the board beyond their values in \cite{Alshowkan2021qlan}: $\cF_{AB}=0.938$, compared to 0.75 before; $\cF_{BC}=0.91$, compared to 0.69; and $\cF_{CA}=0.971$, compared to 0.90. The total ebit rates are slightly lower than before. The cause of this is unknown, but we suspect it derives from extra loss from manual optimization of the setup, such as fiber connections and free-space alignment. These results provide direct confirmation of the substantial performance enhancements possible with WR timing synchronization in the context of quantum networking. Although slightly more expensive than our GPS system~\cite{Alshowkan2021qlan}, WR attains orders of magnitude lower jitter, all within a turn-key system that presents the same outputs to network hardware as GPS---i.e., 1~Hz time pulses and a 10~MHz reference.

\begin{table}[b!]
	\centering
	\caption{Link data for the allocated bandwidth.}
	\label{table_result}
	    \begin{tabular}{c|c|ccc}
		\hline 
		\hline 
		\textbf{Link} & \textbf{Ch.} & \textbf{Fidelity} & $\bm{E_\mathcal{N}}$ \textbf{[ebits]} & $\bm{R_E}$ \textbf{[ebits/s]} 		\\ \hline
		A--B  &  3 	  &  $0.938 \pm 0.008$ & $0.94  \pm 0.02$   & $47   \pm 1$   \\
		B--C  &  1--2 &  $0.91  \pm 0.01$  & $0.91  \pm 0.0.03$ & $19.6 \pm 0.6$ \\ 
		C--A  &  8    &  $0.971 \pm 0.004$ & $0.970 \pm 0.009$  & $145  \pm 1$   \\
		\hline
		\hline
	\end{tabular}
\end{table}

\section{Discussion}
The current configuration is highly scalable. Besides the single-photon detectors, our current WR switch is able to time synchronize 19 nodes and can be scaled by connecting additional switches and receivers. Similarly, our 9-output WSS can be expanded to more users with a commercially available 20-output WSS or by nesting multiple WSSs as described in Ref.~\cite{Alshowkan2021qlan}. Future advances in lightwave technology and WSS production with more outputs will enable us to deploy a QLAN subnet similar in size to the classical network counterpart: 254 networked nodes. Admittedly, attaining such a high number of nodes would benefit from more efficient optical fiber utilization than implemented here, where each user receives three separate strands: one for the entangled photon, another for WR, and a third for the QKD signal. In principle, wavelength multiplexing could be leveraged to combine all signals into a single fiber%
, although further research on crosstalk effects and mitigation will be required to ensure performance is maintained.
As an additional way to reduce the resource burden of traditional pair-wise QKD connections between all network users,  quantum secret sharing (QSS) techniques~\cite{Grice2015two, Kogias2017unconditional, Grice2019quantum, Williams2019quantum, Richter2021agile} based on a full-mesh topology could be considered in lieu of dedicated point-to-point QKD systems for securing future QLANs. 

For the majority of quantum network applications, we suspect that the $\sim$10~ps WR standard deviations observed here [Fig.~\ref{fig_jitter_multi}(a)]---and even lower values possible with a system consisting entirely of components with low-jitter daughterboards~\cite{Rizzi2018}---should prove more than sufficient for node synchronization. Indeed, in our case, we are limited much more strongly by detector and FPGA jitter, making further improvement to clock synchronization superfluous. Nonetheless, it is possible one might require even tighter synchronization for a specific quantum system, in which case optical time transfer systems could be explored as WR alternatives, for which femtosecond-level jitters have been demonstrated~\cite{Foreman2007, Sliwczynski2010}.

From a cybersecurity perspective, we note that our 40~minutes key update period---selected to avoid UDP data loss during extended measurements---vastly underutilizes the capabilities of our QKD system, which produces approximately six new 256-bit keys every second (see Fig.~\ref{fig_key_rate_error}), so that security can be further improved if desired. And while quantum-derived keys secure all classical data communications in our QLAN tested, the WR timing signals may still be vulnerable to delay asymmetry attacks such as those demonstrated against PTP~\cite{Annessi2018}. Countermeasures against such attacks represent an ongoing area of research; nevertheless, we emphasize that the primary consequence of losing synchronization in our case would be a precipitous drop in coincidences---a denial-of-service-type attack that would not expose any secured data to eavesdropping or tampering.

\section*{Acknowledgments}
We thank T. Jones and M. Lipi\'{n}ski for helpful discussions on White Rabbit. This work was performed in part at Oak Ridge National Laboratory, operated by UT-Battelle for the U.S. Department of Energy under contract no. DE-AC05-00OR22725. Funding was provided by the U.S. Department of Energy, Office of Science, Office of Advanced Scientific Computing Research, through the Early Career Research Program and Transparent Optical Quantum Networks for Distributed Science Program (Field Work Proposals ERKJ353 and ERKJ355). Optimization of the Charlie node was supported by the U.S. Department of Energy, Office of Science, Basic Energy Sciences, Materials Sciences and Engineering Division (Field Work Proposal ERKCK51). Postdoctoral research support was provided by the Intelligence Community Postdoctoral Research Fellowship Program at the Oak Ridge National Laboratory, administered by Oak Ridge Institute for Science and Education through an interagency agreement between the U.S. Department of Energy and the Office of the Director of National Intelligence.

%

\end{document}